\documentstyle[12pt,aaspp4]{article}

\slugcomment{Submitted to {\it The Astrophysical Journal}}
\lefthead{Morales et al.}
\righthead{FUV flux of $\alpha$ Vir}

\begin{document}

\title
{Far Ultraviolet Absolute Flux of $\alpha$ Virginis\footnote{Based on the 
development and utilization of the Espectr\'ografo
Ultravioleta de Radiaci\'on Difusa, a collaboration of the Spanish Instituto 
Nacional de T\'{e}cnica Aeroespacial and the Center for EUV Astrophysics, 
University of California, Berkeley}}
\author{Carmen Morales, Joaqu\'{\i}n Trapero\altaffilmark{2}, 
Jos\'{e} F. G\'{o}mez,
\'{A}lvaro Gim\'{e}nez\altaffilmark{3}, Ver\'{o}nica Orozco} 
\affil{Laboratorio de Astrof\'{\i}sica Espacial y F\'{\i}sica
  Fundamental, INTA, 
  Apdo. Correos 50727, E-28080 Madrid, Spain}
\author{Stuart Bowyer, Jerry Edelstein, Eric Korpela, Michael Lampton,
Jeff Cobb}
\affil{Space Sciences Laboratory, University of California,
Berkeley, CA 94720--7304, USA}

\altaffiltext{2}{Present address: Universidad SEK, Cardenal
Z\'{u}\~{n}iga s/n, Segovia, Spain}
\altaffiltext{3}{Also at Instituto de Astrof\'{\i}sica de Andaluc\'{\i}a,
CSIC, Apdo Correos 3004, E-18080, Granada, Spain}

\begin{abstract}

We present the far ultraviolet spectrum of $\alpha$ Virginis taken with EURD
spectrograph on-board MINISAT-01. 
The spectral range covered is from $\sim 900$ to 1080
\AA\ with 5 \AA\ spectral resolution.  
We have fitted Kurucz models to IUE 
spectra of $\alpha$ Vir and compared the extension of the model to our 
wavelengths 
with EURD data. This comparison shows that EURD fluxes are consistent with the
prediction of the model within $\sim 20-30$ \%,
depending on the reddening assumed.
EURD fluxes are consistent with Voyager 
observations but are $\sim$60\% higher than most previous rocket
observations of $\alpha$ Vir.

\end{abstract}


\section{Introduction}

Observations of OB stars in the far-ultraviolet 
(FUV), below 1100 \AA, are important to address several issues, 
such as the interaction of these stars with their surrounding ISM, or
the determination of interstellar dust size. 
These observations can also serve as a test of stellar atmosphere
models and to set spectrophotometric standards in the FUV, which are brighter
and more abundant than white dwarfs, for future space observatories.
 
There are very few FUV spectroscopic observations of stars.
Since the first rocket observations of stellar emission in the FUV
(Brune, Mount, \& Feldman 1979\markcite{bru79}), several attempts have been made 
to determine the spectral energy 
distribution of stars of different spectral types in this wavelength range. In early 
work rockets were used to study stars in the FUV, and there were some spacecraft 
observations by Voyager (Broadfoot et al. 1977\markcite{bro77}) and Copernicus
(Rogerson et al. 1973\markcite{rog73}) spacecraft observations. More
recently, three instruments  
with FUV spectrometers, HUT (Davidsen 1990\markcite{dav90}), ORFEUS
(Hurwitz \& Bowyer 1991\markcite{hur91}) and UVSTAR
(Stalio et al. 1993\markcite{sta93}), have flown on-board the Space Shuttle. 

Absolute calibration in this wavelength range is difficult due to the 
absence of primary
or secondary standards, and indirect calibrations were applied in all
cases. The consequence is that large flux differences, of up to a factor of ten at
certain wavelengths, are found between observations of the same star
taken with different instruments, and none of 
them agrees with Kurucz model atmospheres below 1000 \AA. Voyager
(Ch\'avez, Stalio, \& Holberg
1995\markcite{cha95}) and HUT (Buss, Kruk, \& Ferguson 1995\markcite{bus95}) observations 
report fluxes higher than the model predictions below 1200 \AA. On the other hand,
rocket observations (Brune et al. 1979\markcite{bru79}; Carruthers,
Heckathorn, \& Opal 1981\markcite{car81}; Woods, Feldman, \& Bruner 
1985\markcite{woo85}; 
Cook, Cash, \& Snow 1989\markcite{coo89}) gave substantially lower fluxes than those from Voyager, and also 
lower than the fluxes predicted by Kurucz model atmospheres. 

$\alpha$ Vir is one of the most studied stars in the FUV. It has 
been observed by Copernicus (York \& Kinahan 1979\markcite{yor79}), Voyager 1 and 2 (Holberg et al. 1982\markcite{hol82}), 
and by rockets (Brune et al. 1979\markcite{bru79}; Cook et
al. 1989\markcite{coo89}, and Wilkinson et  al. 1995\markcite{wil95}). 

The fundamental parameters of $\alpha$ Vir are well known.
Spica is a double-lined spectroscopic binary with an ellipsoidal variation of
0.03 mag. due to tidal distortion with 4-day orbital period, 
superposed to $\beta$ Cephei-type pulsations of the primary, 
of 0.016 mag. amplitude and a period of $0\fd1738$ (Shobbrook et
al. 1969\markcite{sho69}).  

The radius of the primary, determined interferometrically by Herbison-Evans et
al. (1971\markcite{her71}), is $R_{1}=8.1\pm0.5$ $R_\odot$ for a distance of $84\pm4$ pc,
and its spectral type is B1V. 
The secondary component of $\alpha$ Vir is probably a B4V star, for which Popper (1980\markcite{pop80})
assumed a $(B-V)=-0.18$, leading to a radius of $R_{2}=4.16\pm1.17$ $R_\odot$ which 
is consistent with the mass ($10.9\pm0.9$ $M_\odot$) deduced by Herbison-Evans
et al. (1971\markcite{her71}). Its parallax has been measured by Hipparcos (12.44$\pm$ 0.86 mas, 
Perryman et al. 1997\markcite{per97}). Corrections for the radii of
the component stars of $\alpha$  
Vir due to this new value of the distance lead to R$_{1}$=7.78 and R$_{2}$=3.99.

Integrated photometry for the two components from the literature as extracted from the
SIMBAD database indicate $V = 0.98$, $B-V = -0.235$, and $U-B = -0.94$ for Johnson
photometry and $b-y = -0.114$, m$_{1}$ = 0.080, and c$_{1}$ = 0.018 for Stromgren uvby
photometry. Crawford H$\beta$ photometry gives $\beta$ = 2.607.

In this paper we present the spectrum of $\alpha$ Vir obtained with EURD 
(Espectr\'ografo Ultravioleta extremo para la Radiaci\'on Difusa)
on-board MINISAT-01. We compare our observations with 
previous ones and with Kurucz models (ATLAS9, Kurucz 1993\markcite{kur93}). Our observations 
allowed us to obtain a flux calibrated spectrum of $\alpha$ Vir with the best 
signal-to-noise ratio and spectral resolution to date in the FUV range.

With this observations we intend to determine the absolute flux of $\alpha$ Vir in the
FUV, for which discrepancies have been found in previous works.

\section{Observations and data reduction}

EURD was launched on April 1997 on-board the Spanish satellite MINISAT-01, 
which has a retrograde orbit of 151$^\circ$ inclination and an altitude of
600 km. Details on the mission can be found in Morales et
al. (1998\markcite{mor98}).

EURD is a spectrograph specially designed to observe diffuse radiation in the
wavelength range from 350 to 1100 \AA. It observes in the anti-sun
direction and
during orbital eclipse. A precise description of the instrument and 
its ground calibration can be found in Bowyer, Edelstein, \& Lampton 
(1997\markcite{bow97}). The
detector is a photon 
counter device that produces spectral images, with spatial resolution
capabilities along an axis perpendicular to the direction of spectral dispersion.
The spectral resolution of the instrument is $\sim 5$ \AA.
When a bright, early-type star falls within the $\sim 25^\circ\times 8^\circ$ field of
view, it shows up as emission longward of $\sim 912$ \AA, on a
finite area along the spatial dimension. Given the mission pointing constraints, only 
stars within $\sim 13^\circ$ from the ecliptic
can be observed.

Observations of $\alpha$ Vir presented in this paper were taken from
1998 April 13 to April 19, and 1999 April 2 to April 23. 
During the data reduction process, we tracked the position of the
stellar emission on the detector for every second, extracted the
photons detected within 10\farcm 5 from the emission maximum 
(to include all the stellar
emission gathered during one second), and subtract a background from
an area of the detector close to where the stellar emission lies.
Photon counts are then corrected by the efficiency of the detector as a function of
incidence angle.
The accumulated photon counts are converted to fluxes by applying the 
in-flight calibration performed with simultaneous observations of the full Moon 
with EUVE and EURD (Edelstein et al., in preparation). 

\section{Results and discussion}

\subsection{The Far-UV spectrum of $\alpha$ Vir}

\begin{figure}
\epsscale{0.8}
\plotone{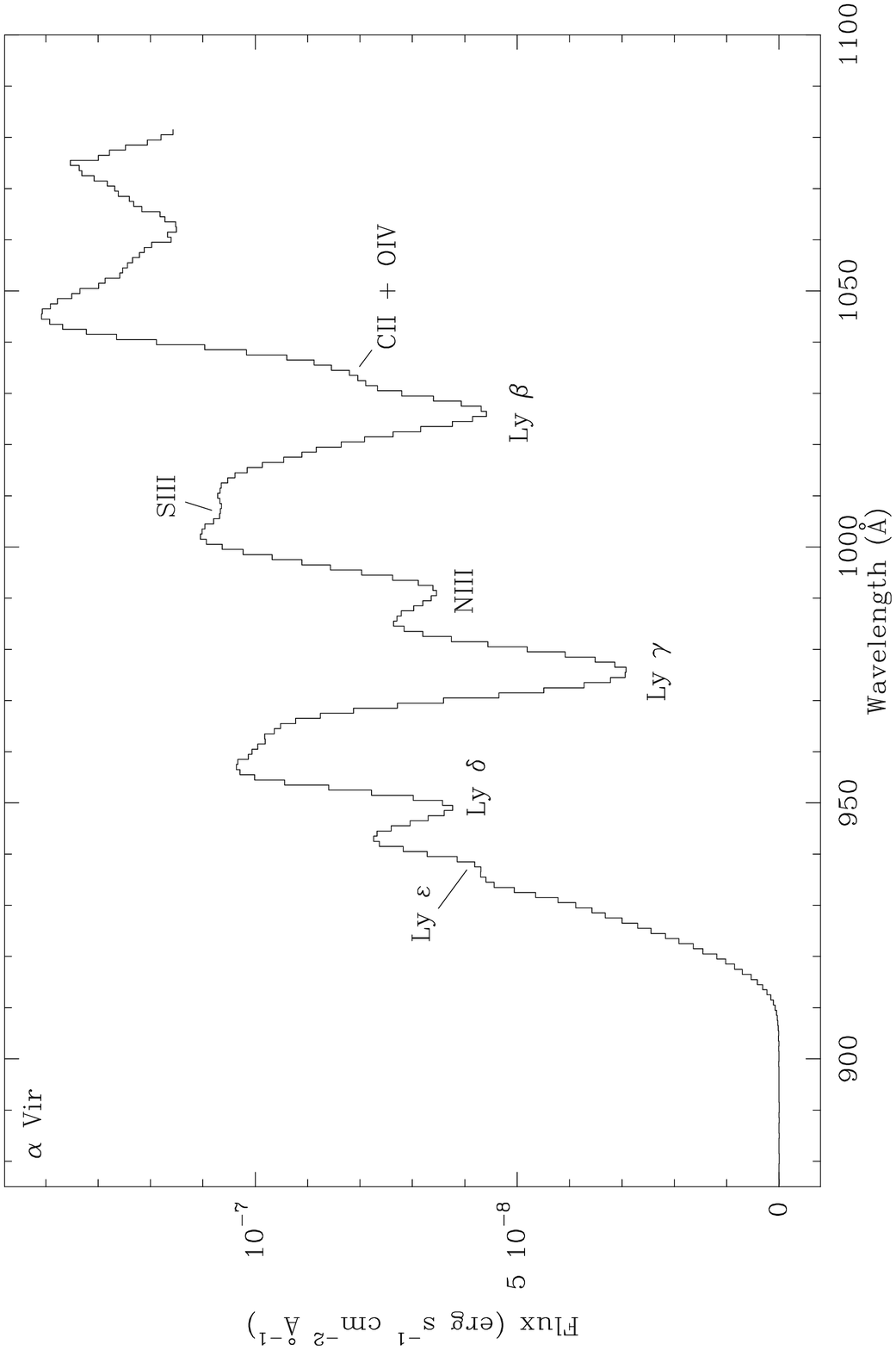}
\caption{\label{whole} Spectrum of $\alpha$ Vir as
observed by EURD. Labels indicate the observed absorption features.}
\end{figure}

Fig. \ref{whole} shows the spectrum of $\alpha$ Vir , with a 
total integration time of $1.06\times10^5$ s, and covering the
wavelengths $\lambda < 1080$ \AA. The noise level of the spectrum 
is $\sigma \simeq 3.3\times10^{-11}$ erg$^{-1}$ cm$^{-2}$ \AA$^{-1}$ and the signal 
to noise ratio is $> 3000$. Some absorption features are obvious in
the spectrum: the 
Lyman series of hydrogen at 937, 949, 972 and 1026 \AA, \ion{N}{3} at 989 \AA, 
\ion{S}{3} at 1012 \AA, \ion{C}{2} and \ion{O}{4} blended at 1037 \AA,
and an unidentified feature at $\sim$1063 \AA\ (see
York \& Kinahan 1979\markcite{yor79}).

\subsection{Comparison with model atmospheres}

Atlas 9 Kurucz models properly reproduce the UV and 
optical spectra of B stars (Malagnini et al. 1985\markcite{mal85}; Fitzpatrick \& Massa 1998\markcite{fit98};
Ch\'avez et al. 1995\markcite{cha95}). For wavelengths below 1200 \AA, 
Holberg et al. (1982\markcite{hol82}) reported for $\alpha$ Vir a flux excess with respect to Kurucz 
models which is within the range of uncertainty of the reddening. They used a model 
atmosphere combination of 24500 K and 17000 K, and a reddening of $E(B-V)=0.02$.
In this wavelength range Buss et al. (1995\markcite{bus95}) also found a flux excess of $\sim 5 \%$ at 
1000 \AA\ with respect to Kurucz models in their sample of galactic OB stars.
 
In this work, we first compared Kurucz models with IUE spectra of $\alpha$ Vir and 
then we checked if the model extension to our wavelengths properly fits the EURD 
spectra. The FUV flux recorded by EURD is the combined flux of the 
two components of 
$\alpha$ Vir binary system. 
Therefore it is necessary to build a combined Kurucz model corresponding to the 
$\alpha$ Vir system. For this purpose we have adopted the effective temperatures
given by Popper (1980\markcite{pop80}), T$_{1}$ = 24500 and T$_{2}$ = 17200. We computed the model 
used for the fit as a combination of Kurucz models of these temperatures (taking into 
account the radii of the two components), and gravity $\log g = 3.69$.

Spica IUE spectra (SWP33091HL and LWR13650HL) have been selected from the 
INES database. To create a single spectrum, we used the SWP spectrum shortward of
1940 \AA, and the LWP one for longer wavelengths. 
The IUE data were 
degraded to the 10 \AA\ spectral resolution of the models. 
Correction for interstellar H{\sc I} and H$_2$ absorption are 
negligible, given the low hydrogen column density (Fruscione et
al. 1994\markcite{fru94})  in front 
of this relatively close star.

\begin{figure}
\epsscale{0.8}
\plotone{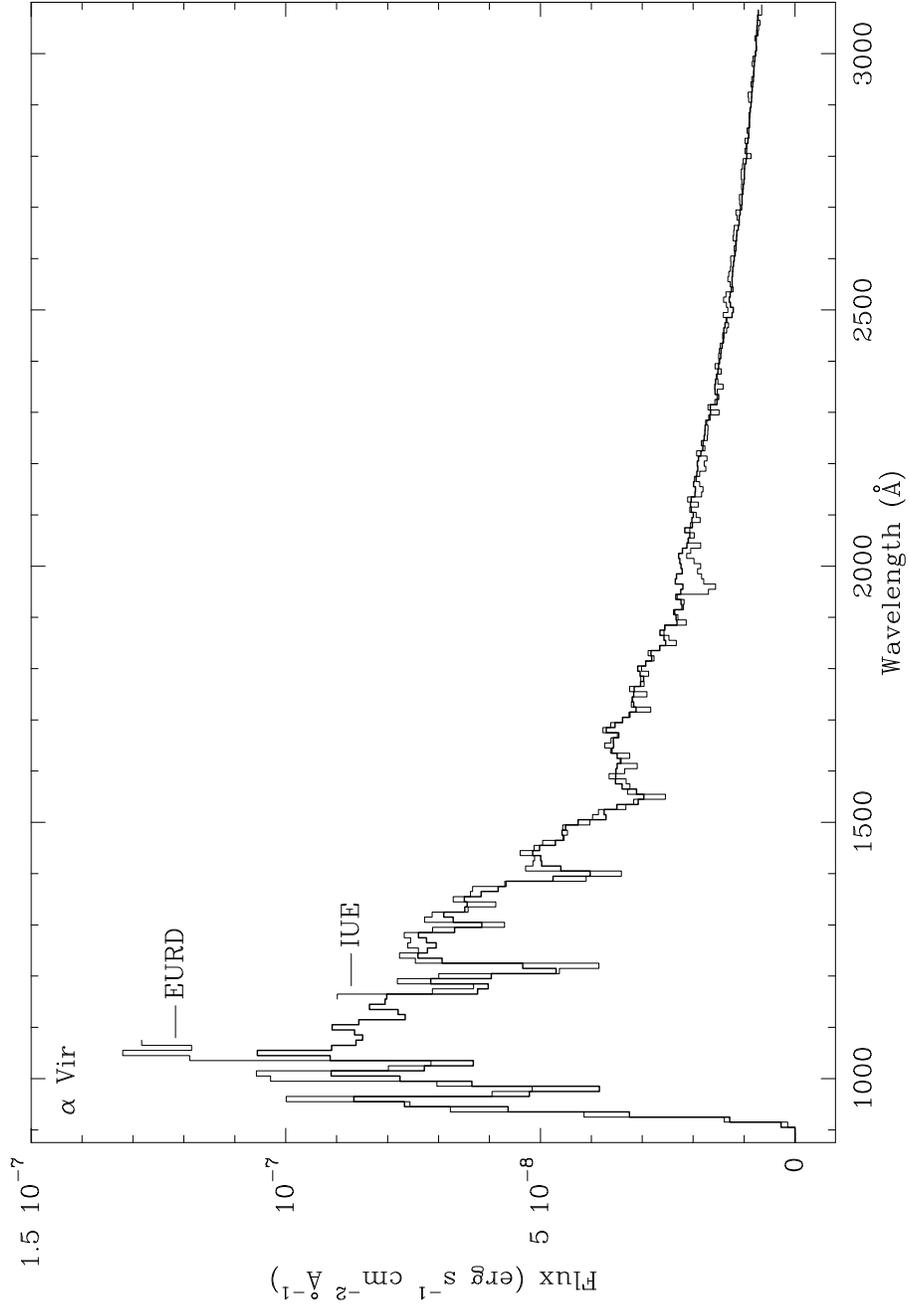}
\caption{\label{withiue} Spectra of $\alpha$ Vir as
observed by EURD ($\lambda < $ 1070 \AA) and IUE ($\lambda > 1160$ \AA),
superposed on the Kurucz model (heavy line) scaled to IUE data. Both
EURD and IUE spectra have been binned down to match
the spectral resolution of the model (10 \AA).}
\end{figure}

We normalized the resultant Kurucz model to the IUE spectrum. 
Adjusting to IUE
observations we avoid the uncertainties possibly present in the 
determination of
the angular radius. Fig. \ref{eurdkurucz} shows the Kurucz model scaled 
to IUE, together with the EURD spectrum applying no reddening
correction, binned down to match the spectral resolution of the models.
The flux of the observed spectrum is $\sim20$\%  higher than the model
flux. Fig. \ref{withiue} shows the comparison of the Kurucz model with
both EURD and IUE spectra.

We have also checked how a possible color excess could affect our results.
The reddening correction is crucial at this wavelength range. Even 
though $\alpha$ Vir has a very low optical reddening, at FUV wavelengths the 
extinction rises very steeply and must be corrected very 
carefully for an absolute flux determination. For wavelengths longer
than 1200 \AA\ the 
interstellar extinction has been well studied (Savage \& Mathis 1979\markcite{sav79}; Seaton
1979\markcite{sea79}; Cardelli,
Clayton and Mathis 1989\markcite{car79}; Fitzpatrick 
\& Massa 1990\markcite{fit90}). However, for wavelengths below 1200 \AA, few attempts
have been done 
to determine an extinction law (Longo et al. 1989\markcite{lon89};
Snow, Allen, \& Polidan 1990\markcite{sno90}; 
Buss et al. 1994\markcite{bus94}). The extinction laws obtained agree in shape with 
the extrapolation of Cardelli et al. (1989\markcite{car79}) law, but their absolute values are very
dependent of the 
value of the ratio of total to selective extinction, $R_v$, in the
star direction. 

We have used the extrapolation of the average extinction law of
Cardelli et al. (1989\markcite{car79}),
applying $R_v=3.1$ for the diffuse interstellar medium, which is the
average of the two more recent and accurate determinations of mean
values of $R_v$ (3.08, He et al. 1995\markcite{he95}; 3.12, Whittet \& van Breda 1980\markcite{whi80}). 
We have followed three different methods to obtain $E(B-V)$: column
density of hydrogen, Str\"{o}mgren uvby photometry, and Johnson UBV
photometry. The highest value of color excess for $\alpha$ Vir 
($E(B-V)=0.02$) was obtained using Johnson UBV
photometry and the intrinsic colors of 
Schmidt-Kaler (1982\markcite{sch82}), taking the
observed $(\bv)=-0.235\pm0.08$ as derived from the recent ground UBV
photometry carried out for the Hipparcos Mission (Perryman et
al. 1997\markcite{per97}). This color excess of 0.02 
is the same used by Holberg et
al. (1982\markcite{hol82}).
Using this value to deredden IUE and EURD spectra, and scaling the
corresponding Kurucz model to the new IUE flux, we find that EURD
fluxes in this case would be in excess by 30\% with respect to the
model expectations. Since the flux excess using $E(B-V)= 0$ was 20\%,  we
can see that, in this case, the redenning correction is not critical in order 
to compare our results with the models.

We conclude that the data are reasonably consistent with 
Kurucz models both in flux and spectral shape. 
A fit to within $5\%$, would be obtained 
by increasing the adopted temperature of the primary by 600 K and 
assuming no reddening.

\subsection{Comparison with previous observations}

There have been very few observations of stellar spectra in the
FUV. Among them, $\alpha$ Vir is one of the most studied stars in this
range (Brune et al. 1979\markcite{bru79}; Holberg et
al. 1982\markcite{hol82}; Cook et al. 1989\markcite{coo89}; Wilkinson et al.
1995\markcite{wil95}).
An indication of the difficulties involved in this study is that
there exist significant differences in the flux derived for this star
by different authors. Cook et al. (1989\markcite{bru89}) and Brune et
al. (1979\markcite{bru79}) spectra are mutually 
consistent, but they are lower than those taken by Voyager (Holberg 
et al. 1982\markcite{hol82}) and recently by Wilkinson et al. 
(1995\markcite{wil95}) with a sounding rocket.

\begin{figure}
\epsscale{0.8}
\plotone{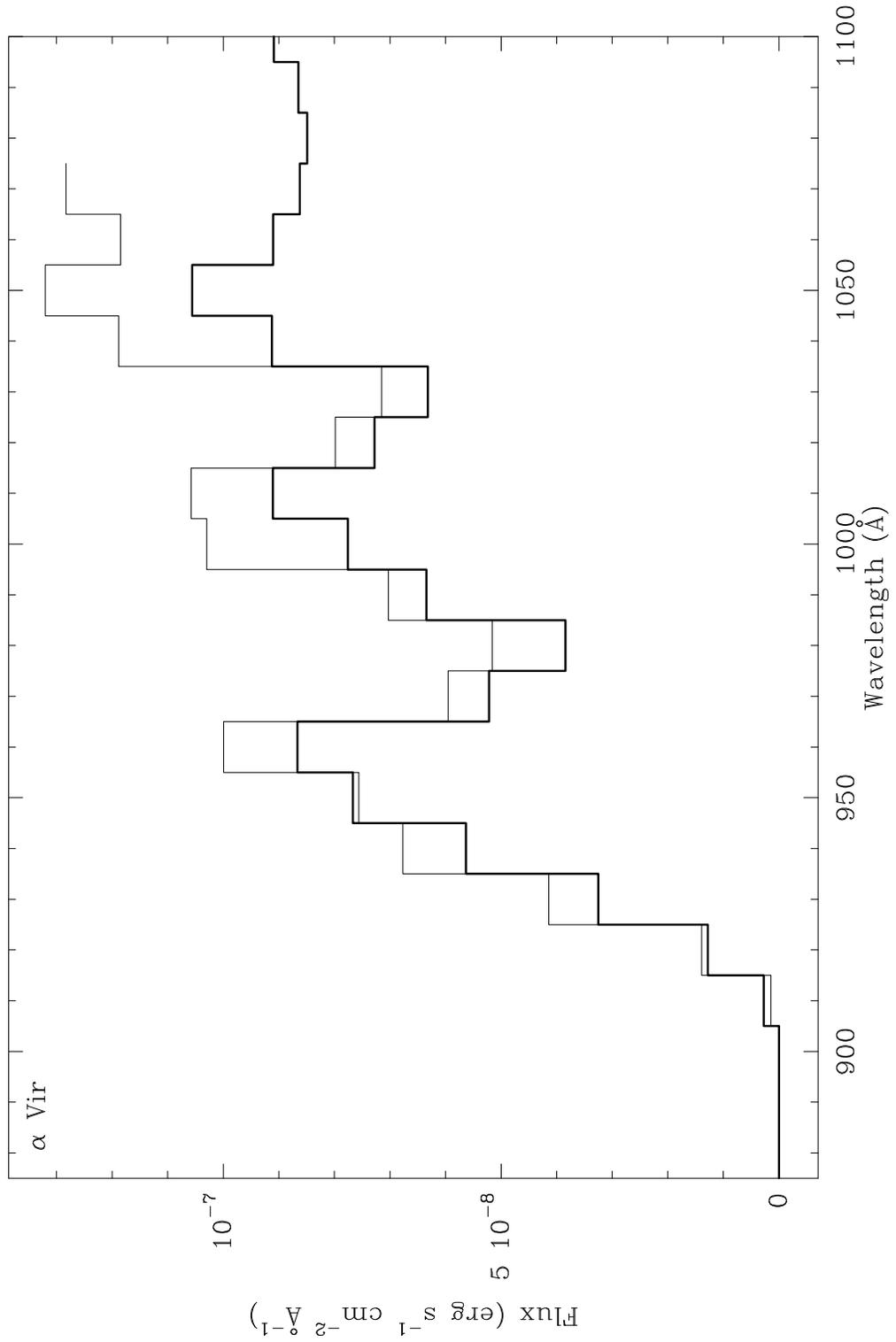}
\caption{\label{eurdkurucz} Spectrum of $\alpha$ Vir
observed by EURD, superposed on the corresponding Kurucz  
model (heavy line). The EURD spectrum has been binned down to match
the spectral resolution of the model (10 \AA).}
\end{figure}

\begin{figure}
\epsscale{0.8}
\plotone{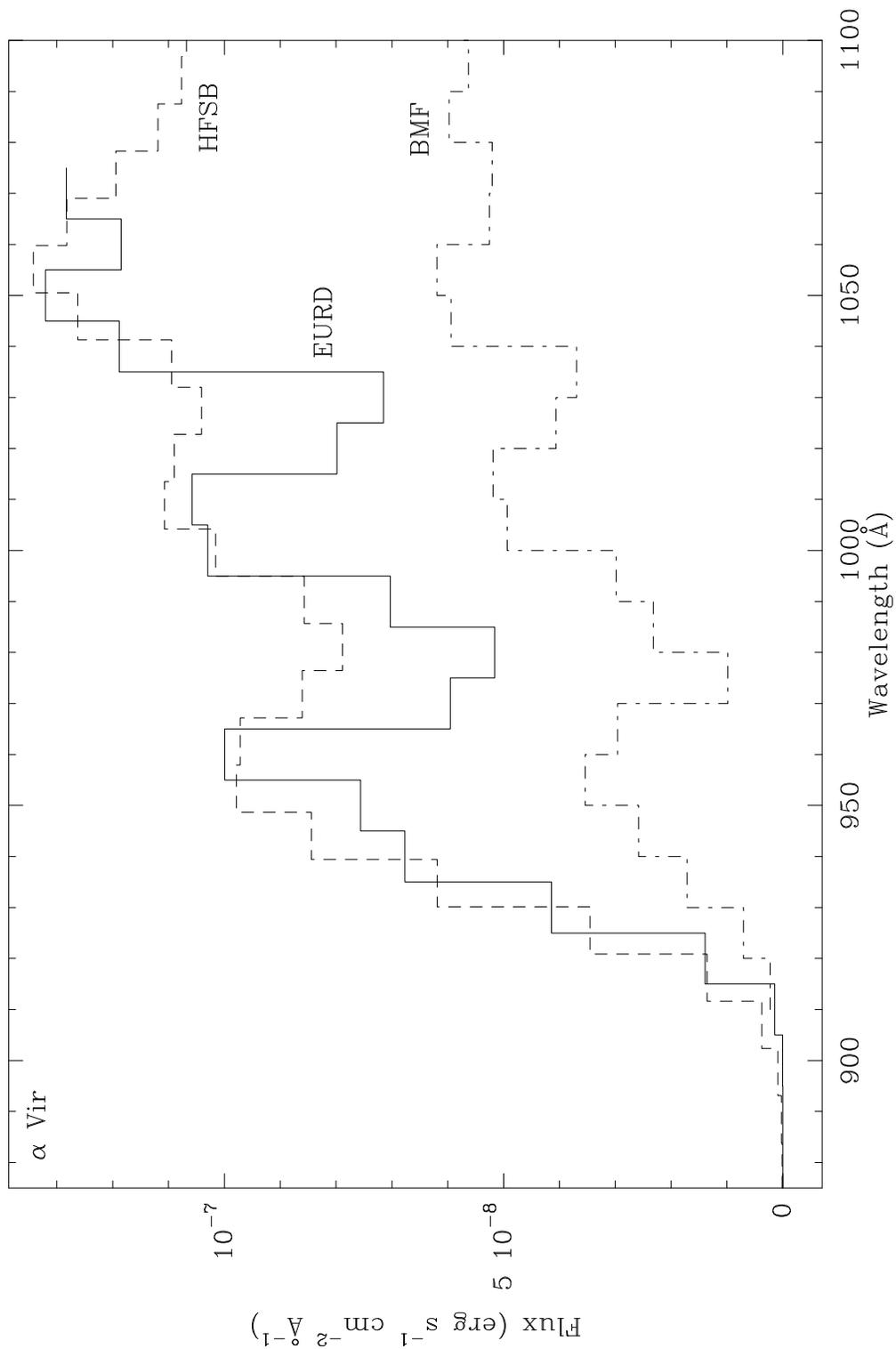}
\caption{\label{compare} Spectrum of $\alpha$ Vir observed by EURD (solid
line), by Holberg et al. (1982) (HFSH, dashed line) and by Brune et al. (1979)
(BMF, dashed-dotted line). The EURD spectrum has been binned down to a
spectral resolution of 10 \AA.}
\end{figure}

Fig. \ref{compare} shows the EURD spectrum compared with previous
observations of $\alpha$ Vir. The fluxes we derive are similar to
those obtained by Holberg et al. (1982\markcite{hol82}). By degrading 
our spectral resolution 
to match that of Holberg et al. (1982\markcite{hol82}) observations, 
we see only a small discrepancy 
in the Lyman lines of hydrogen, whose absorption is deeper in the EURD
spectrum. This can be due to residual atmospheric extinction 
in our data, which is not present in Voyager observations. Note,
however, that the depth of the Lyman absorption lines in EURD data are
more 
consistent with the prediction of the Kurucz models (Fig.\ \ref{eurdkurucz}). 

As for the flux discrepancy between previous observations, we suggest
that the fluxes obtained by EURD and Voyager better represent the real
far-UV flux of $\alpha$ Vir, while the group of observations that
provide lower fluxes (Cook et al. 1989\markcite{bru89}; Brune et
al. 1979\markcite{bru79}) seems to underestimate it.
This suggestion is supported by the better consistency between our
data and those of IUE, since the EURD spectrum is close to
the Kurucz model that better reproduce IUE data (Fig.\ \ref{eurdkurucz}).
The situation could switch in favor of the lower-flux observations
if Kurucz models proved to overestimate the fluxes in this wavelength
region.

\section{Conclusions}

We present new $\alpha$ Vir observations below 1080 \AA\ with improved spectral 
resolution and signal to noise ratio, taken with the EURD spectrograph on-board MINISAT
01. We compared the EURD $\alpha$ Vir spectrum with a Kurucz model atmosphere computed
with the best values of the temperature, distance and radius of the components of 
$\alpha$ Vir binary system. This
comparison shows that models 
are in reasonable agreement with the flux measured by EURD, being 20-30\% higher 
than the models.

Our results support Voyager fluxes of Holberg et al. (1982),
rather than the lower fluxes given by rocket observations.

\acknowledgments

This research has made use of the SIMBAD database, operated at CDS, Strasbourg,
France, and of the International Ultraviolet Explorer data retrieved from the INES
Archive.
The development of this instrument has been partially supported by INTA grant IGE
490056. C.M. and J.T. acknowledge support by 
DGCYT grant PB94-0007. JFG is supported 
in part by DGICYT grant PB95-0066 and by Junta de Andaluc\'{\i}a (Spain). Support for
the publication of this paper has been provided by INTA. 

The UCB authors wish to thank Yumi Odama for help in the data processing. Partial
support for the development of the EURD instrument was provided by NASA grant NGR
05-003-450. When NASA funds were withdrawn by Ed Weiler, the instrument was completed
with funds provided by S. Bowyer. The UCB analysis and interpretation work is carried
out through the volunteer efforts of the authors.

\clearpage

\end{document}